# YouTube and political communication – Macedonian case

MSc. Sali Emruli[1], Mr. Tahir Zejneli[2], MSc. Florin Agai[3]

**Faculty of Organization and Informatics, University of Zagreb,**
Varaždin, 42000, Croatia

**Faculty of Communication Sciences and Technologies, SEE-University,**
Tetovo, 1200, Macedonia

**Faculty of Electronics, Telecommunication and Informatics, University of Pristina,**
Pristina, Kosovo

## Abstract

Analysis how to use Internet influence to the process of political communication, marketing and the management of public relations, what kind of online communication methods are used by political parties, and to assess satisfaction, means of communication and the services they provide to their party's voters (people) and other interest groups and whether social networks can affect the political and economic changes in the state, and the political power of one party.

## 2. Introduction

The analysis will be done in a way that will create a list of largest political parties in the Republic of Macedonia, their communications infrastructure through ICT (website), content and the manner in which they placed their information and receive feedback from voters.

Internet, social networking, Web 2.0, Facebook, YouTube, blog ... All these are relatively new word in the political vocabulary, new concepts, new media and new opportunities for the transmission of ideas and messages are not enough channels used to communicate with the public. Although the practice of using the Internet in local political advertising goes back to the nineties, only in recent years the advent of new tools and social networks demonstrates true strength of this medium.

Besides direct access to the public, political ideas, it provides full force confrontation, but also provides a relatively convenient ground for review of public attitudes, research and development of certain ideas. Using such a change in social communication, transmission of political messages through the transition from traditional forms of communication and finding new paths to the recipients.

Professional and political public for years following the development of the Internet as a medium, but he showed the greatest strength in the last U.S. presidential election.

Political power depends on the satisfaction of the people towards a particular party and party connections with other parties or organizations. Well-developed social network provides further prestige and power of the party and its direct channel of communication with voters and other influential interests groups.

## 3. Internet and countries in transition, development and application

What are the position and the impact of the less developed countries and countries in transition on the development and application of the Internet?

Countries in transition depend on how they can establish cooperation with the developed centers to remain or not to remain marginal areas. The problem is the fact that less developed countries, specifically countries in transition differently perceive, and thus they apply Internet technologies and new ways of communicating.

In The information-communication sphere Internet is an incarnation of globalization.

As globalization breaks down communication sovereignty of the state, to a small state and regions in transition, especially the states founded by the collapse of previous socialist countries creates an additional problem - the problem of relationship between global and national identity and the deterioration of their difficult and recently acquired sovereignty.

However, rare are governments in less developed countries that without any resistance would agree to such changes, but also to those in the communication process. Internet and network connection, in an ideal sense, it creates an open communication space and the possibility of an open and democratic pluralism. Openness rather than closure, decentralization instead of centralization, the availability of information rather than secrecy, multilateral (many-side) rather than the unidirectional transfer of information and cooperation instead of antagonism are the key ideas promoted by Internet.

Can local and national governments (especially small and countries in transition) respond to these challenges, where there is still a relationship of distrust and tension between the government and the media, and instead of publics in function there is the secrecy of work, which justice is often the security, political and security reasons. The relationship of mistrust is even more to the Internet as a free zone "with no guards at the gates" ("gatekeeper-free").

Which are, therefore, the basic conditions involving "small" states into "high" global information and communication flows? As we have already said, there are three basic tasks that national governments must meet if they want to follow the global (ICT) trends:

Transformation of Consciousness (government institutions must be in accordance with the principles of openness and cooperation, typical Internet),
(faster) adoption of technology, and
Adaptation of the system (closed to open).

## 4. Social networking importance on politics

According to Danah Boyd and Nicole Ellison, social networking sites can be defined as web-based services that allow individuals to:

*(1) Construct a public or semi-public profile within a bounded system,*
*(2) Articulate a list of other users with whom they share a connection, and*
*(3) View and traverse their list of connections and those made by others within the system."*
According to this definition, the first social networking website was launched in 1997.

There are four main reasons for social networks playing more prominent parts in political advertising strategies:

**1. Large reach:** Recently, the top three social networking sites, such as Facebook, Twitter and YouTube, had roughly 800 000 unique visitors combined in the Macedonia alone. This accounted for roughly one-thirds of the total Macedonia population online and nearly half of the total Macedonian population. Additionally, internet users worldwide are spending increasing amounts of time on activities with social connections. By incorporating social networks in their political advertising strategies, Macedonian political parties can reach out to a large user base spread across the country with relative ease.

**2. Cost efficiency:** Political Advertising on social networks is relatively cheap compared to other traditional media: it usually has a similar or expanded reach at much lower costs. In addition, it is possible for political parties to generate free publicity through creative political advertising techniques. There have been a number of successful political marketing campaigns on YouTube and Facebook in recent years.

**3. Targeted advertising:** Political Advertisers have access to a great deal of information about users and their interests, allowing them to customize and target their ads to a degree not yet seen in any other advertising medium.

**4. Time spent online:** People are spending increasing amounts of time online, especially on social networking websites, at the expense of traditional advertising media such as television and newspapers. This can be viewed as a result of many of the traditional functions – news, television shows – of the old advertising media being shifted online to social networks such as Facebook and YouTube. As a result, political party advertisers are increasingly looking for ways to reach out to voters who are spending more and more time online.

However, despite numerous efforts of political parties to reach out to voters, they are becoming increasingly disenchanted by the lack of results.

## 5. Political Parties in Republic of Macedonia
## 5.1. Overview of the political system

Macedonia is a Republic having multi-party parliamentary democracy and a political system with strict division into legislative, executive and judicial branches. From 1945 Macedonia had been a sovereign Republic within Federal Yugoslavia and on September 8, 1991, following the referendum of its citizens, Macedonia was proclaimed a sovereign and independent state. The Constitution of the Republic of Macedonia was adopted on November 17, 1991, by the first multiparty parliament. The basic intention was to constitute Macedonia as a sovereign and independent, civil and democratic state and also to create an institutional framework for the development of parliamentary democracy, guaranteeing human rights, civil liberties and national equality.

The Assembly is the central and most important institution of state authority. According to the Constitution it is a representative body of the citizens and the legislative power of the Republic is vested in it. The Assembly is composed of 120 seats.

The President of the Republic of Macedonia represents the Republic, and is Commander-in-Chief of the Armed Forces of Macedonia. He is elected in general and direct elections, for a term of five years, and two terms at most.

Executive power of the Republic of Macedonia is bicephalous and is divided between the Government and the President of the Republic. The Government is elected by the Assembly of the Republic of Macedonia by a majority vote of the total number of Representatives, and is accountable for its work to the Assembly. The organization and work of the Government is defined by a law on the Government.

In accordance with its constitutional competencies, executive power is vested in the Government of the Republic of Macedonia. It is the highest institution of the state administration and has, among others, the following responsibilities: it p roposes laws, the budget of the Republic and other regulations passed by the Assembly, it determines the policies of execution of laws and other regulations of the Assembly and is responsible for their execution, decides on the recognition of states and governments, establishes diplomatic and consular relations with other states, proposes the Public Prosecutor, proposes the appointment of ambassadors and representatives of the Republic of Macedonia abroad and appoints chiefs of consular offices, and also performs other duties stipulated by the Constitution and law.

In Macedonia there are more political parties participating in the electoral process at national and local level.

### 5.2. Current Structure

Parties of traditional left and right:
Coalition VMRO – DPMNE (63 mandates, right oriented Macedonian party)
Democratic Party of the Albanians (12 mandates, right oriented Albanian party)
Coalition "SONCE" – SDSM (27 mandates left oriented Macedonian party)
Democratic Union for Integration (18 mandates left oriented Albanian party)

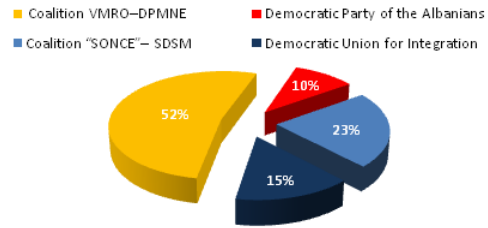

Fig. 1 Current political structure in Republic of Macedonia and Political Parties mandate percentage win in latest parliamentary elections.

## 6. Analysis: Political Parties on Social networks in Macedonia

The data on these parties' Web 2.0 presence was gathered through three types of sources. The first source was the official party web sites (Screenshots of political parties – Appendix), whose front pages were scanned for links to external Web 2.0 pages, as well as Web 2.0 elements built or embedded within the web site itself. This gives a good indication of what was officially approved by the party in question, in the sense of representing the party organization, as well as having a sufficient quality.

Most of the other parties were more cautious and advertised a s ingle element, either Facebook or blogs, although other data indicate they were present elsewhere as well.

| Political Party | Facebook | Twitter | Blogs | YouTube | Flicker | Other | Total | Percentage |
|---|---|---|---|---|---|---|---|---|
| VMRO - DPMNE | 0 | 0 | 0 | 0 | 0 | 1 | 1 | 17% |
| Social - Democratic Union of Macedonia | 1 | 1 | 1 | 1 | 0 | 1 | 5 | 83% |
| Democratic Union for Integration | 0 | 0 | 0 | 0 | 0 | 0 | 0 | 0% |
| Democratic Party of the Albanians | 0 | 0 | 0 | 0 | 0 | 0 | 0 | 0% |
| Liberal - Democratic Party | 1 | 0 | 1 | 0 | 0 | 1 | 2 | 33% |
| New Democracy | 1 | 1 | 0 | 1 | 0 | 1 | 4 | 67% |
| New Social-Democratic Party | 0 | 0 | 0 | 0 | 0 | 0 | 0 | 0% |
| Socialist Party | 1 | 0 | 0 | 0 | 0 | 0 | 1 | 17% |
| Democratic reconstruction of Macedonia | 0 | 0 | 0 | 0 | 0 | 0 | 0 | 0% |
| Democratic union | 0 | 0 | 0 | 0 | 0 | 0 | 0 | 0% |
| Democratic Party of Serbs in Macedonia | 1 | 1 | 0 | 1 | 0 | 1 | 4 | 67% |
| Democratic Party of Turks in Macedonia | 0 | 0 | 0 | 0 | 0 | 0 | 0 | 0% |
| Liberal Party | 0 | 0 | 0 | 0 | 0 | 0 | 0 | 0 |

Table 1: Social networking elements on front page of parties' websites.

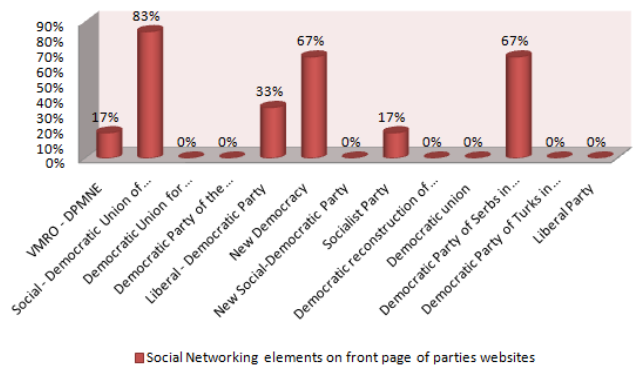

Fig.2 Social networking elements on political parties' websites front page

## 6.1. YouTube
### 6.1.1. Introduction

With the development of YouTube in 2005, and its rising popularity as a campaign tool, this study aims to explore the changing role of the Internet, with particular emphasis on YouTube and its effects on politics. There are examples that demonstrated the power of online video in impacting constituent opinion and election results. In conducting the above analysis, the study determined that the content on YouTube that generates the highest number of page views includes both negative campaign content and informative video clips. As a co nsequence, YouTube has provided important opportunities for independent actors to play a role in the context of a political campaign.

YouTube, an Internet website that hosts a v ast array of video content, was developed for users to upload video, share clips with their friends, and provide a f orum for social networking around video content. When the political world took note of this new medium, they believed that it would provide candidates and constituents with the opportunity to post web video on and about the candidate. To date, the content posted on the site, by both the campaign and voters, ranges from news clips, speeches, interviews, comedy, satire, and negative campaign content. During the elections, we first see user and campaign generated content in a p olitical campaign, the results of which led to the uploading of a l arge and wide range of video, including features both praising and insulting candidates. For example, at the beginning 2008 election, the top three Democratic presidential candidates (John Edwards, Hillary Clinton and Barack Obama) all announced their presidency via online video posted on YouTube.

YouTube and online video represent another way for constituents to collect information about a candidate. Therefore, candidates, no matter what party they are affiliated with, have decided that they can benefit by engaging with the voter through this medium. YouTube marked a turning point in politics on the Internet. Not only is YouTube providing an arena for video content in a way not previously available but information is being uploaded immediately and disseminated to the public faster than ever before. Furthermore, the Internet, particularly YouTube, is offering constituents and campaigns the opportunity to release information that is transparent, effortless in its distribution, and allows for voter participation, unlike TV which has an element of control that online video lacks.

Video sharing sites will become an increasingly important medium for reaching voters as a greater proportion of the electorate logs on. YouTube will influence elections by extending the reach of campaign materials; enabling candidates with little funding to get "face time;" reaching young people through entertaining, visual means; democratizing political information while diminishing candidate's ability to control their message; increasing attention to negative advertising; and putting scandals on the agenda and/or extending their shelf life. YouTube enables candidates, political parties, and interest groups to reach an extended segment of the population with minimal cost. This will likely increase the number of candidates in races, especially the presidential election, where early exposure and momentum are crucial. YouTube allows more candidates to follow in the footsteps of "Howard Dean" and use the Internet to gather "underground" momentum. While it is unlikely that little-known presidential candidates will have a large audience on YouTube, those with some name recognition will benefit from having their ads accessible to the general public without expensive media buys. YouTube is well on its way to become an established part of the parties' communication strategies.

### 6.1.2. Limitations

Limitation was the impossibility of accessing official YouTube video channels of political parties from their front pages because they did not have direct link to their video channels on YouTube, exception is a small number of political parties. Because of these limitations finding of official YouTube video channels of political parties is done with unusual keywords.

### 6.1.3. Methodology

As a result of the above mentioned limitations the research is done in order to find official YouTube video channels of political parties not with the original name nor their abbreviation but with different key words, as for example for Democratic Union for Integration (eng. DUI, alb. BDI) the official YouTube channel of this political party was found with key word bdi2009.

The analysis is based on some attribute of YouTube's as: Date of joining, Number of subscribers, Number of Videos in the channel, Views of the channel, Total number of Video views, number of comments and links with other YouTube accounts.

### 6.1.4. Results

The obtained results shows that from the thirteen political parties analyzed, nine (69%) have official video channels on YouTube while four (31%) have not video channels. The table below also presented official channel links of political parties which has YouTube video channel.

| Political Party | YouTube channels | Links |
|---|---|---|
| VMRO - DPMNE | 1 | http://www.youtube.com/user/vmrodpmnemacedonia |
| Social - Democratic Union of Macedonia | 1 | http://www.youtube.com/user/SDSMakedonija |
| Democratic Union for Integration | 1 | http://www.youtube.com/user/bdi2009 |
| Democratic Party of the Albanians | 1 | http://www.youtube.com/user/gurrapdsh |
| Liberal - Democratic Party | 1 | http://www.youtube.com/user/liberaldemocraticpar |
| New Democracy | 1 | http://www.youtube.com/user/DRMEDIACENTRE |
| New Social-Democratic Party | 1 | http://www.youtube.com/user/nsdp2008 |
| Socialist Party | 0 | N |
| Democratic reconstruction of Macedonia | 1 | http://www.youtube.com/user/dommakedonija |
| Democratic union | 0 | N |
| Democratic Party of Serbs in Macedonia | 1 | http://www.youtube.com/user/dpsmdpsm |
| Democratic Party of Turks in Macedonia | 0 | N |
| Liberal Party | 0 | N |
| TOTAL | 9 | |

Table 2: Thirteen political parties in Republic of Macedonia which we have taken in this research and their YouTube official channel links.

From the results presented in the table can be seen that some political parties have no official video channel on YouTube, but this does not mean that there are not any video clips that are dedicated to the activities of those political parties. Video clips for which we cannot say with accuracy that are posted by the political party are not taken into consideration but are calculated as fans activity rather than as a direct activity of political parties. Video materials posted by different individuals exist in large numbers for any political party.

From the table below can be seen the joining dates of official video channels of political parties on YouTube, also from the dates we can conclude that most of the political parties have created their official video channels during the parliamentary elections which were held in 2008, municipal elections and presidential elections which were held in 2009.

| Political Party | Mandates | Joined | Subscribed | Videos | Channel views | Total upload views | Comments | Linked |
|---|---|---|---|---|---|---|---|---|
| VMRO - DPMNE | 55 | 30.06.2009 | 227 | 597 | 7534 | 80,925 | 119 | 450 |
| Social - Democratic Union of Macedon | 18 | 28.04.2008 | 71 | 190 | 6956 | 110,251 | 5 | 39 |
| Democratic Union for Integration | 18 | 01.03.2009 | 2 | 17 | 391 | 18,026 | N | N |
| Democratic Party of the Albanians | 5 | 04.02.2009 | 7 | 30 | 4110 | 112,289 | N | N |
| Liberal - Democratic Party | 4 | 19.04.2008 | 3 | 1 | 746 | 380 | N | N |
| New Democracy | 4 | 31.03.2010 | 1 | 27 | 232 | 2,609 | N | N |
| New Social-Democratic Party | 3 | 23.04.2008 | 3 | 16 | 122 | 2,283 | N | N |
| Socialist Party | 1 | N | N | N | N | N | N | N |
| Democratic reconstruction of Macedon | 1 | 17.10.2007 | 3 | 11 | 256 | 2,043 | N | N |
| Democratic union | 1 | N | N | N | N | N | N | N |
| Democratic Party of Serbs in Macedon | 1 | 09.05.2010 | N | 34 | 88 | 2,839 | N | N |
| Democratic Party of Turks in Macedon | 1 | N | N | N | N | N | N | N |
| Liberal Party | 1 | N | N | N | N | N | N | N |
| Total | | 9 | 317 | 923 | 20535 | 331,645 | 124 | 489 |

Table 3: Attributes examined in the analysis.

Correlation between the seats won by political parties in recent parliamentary elections and their activity in this social medium exists. By thirteen political parties obtained in the research (political parties which have their official web page and which uses social media) we can conclude that the political parties which have the largest number of mandates won in the last parliamentary election (VMRO-DPMNE) have most pronounced activity in this social media. Starting from the number of subscribers, video material posted on their channels, official channels attendance and viewership of published videos on these channels are in proportion with the mandates received.

From the table we can also conclude that the political parties with dominant number of seats in the Macedonian parliament are more transparent regarding to the possibility of commenting on posted video materials on their official video channels.

Some political parties are using YouTube as medium for the announcement of Press Conference, tribunes, interviews, party rallies and some events without the right to comment those video because of political insults (comments had been disabled by the administrator of the channel), that means the unilateral communication with the electorate, without transparency and unsocial, which is opposite to the nature of social media aimed interactive two-way communication, and condescension of the opposite opinion.

## 7. Conclusion

With the introduction of user generated video content to the web, but more importantly into the political sphere, and the development of YouTube in 2005, the face of Internet politics has become less about what is being said and increasingly focused on how it is being said. It provided a space in which anyone was able to contribute to a political discussion despite the voter's location, political affiliation, opinions and/or thoughts.

YouTube, like the blogs, provided a space for voters and campaign members to upload video clips, regardless of the length, to the site, as a housing location for web videos and a place for all constituents to view the posted content. This is the first time, in the history of political campaigns, that we saw online video influence a campaign, both during the campaigning process and within the election results. Most importantly, online video and YouTube opens up the democratic process as it provides a forum and means for two way communication, not solely the traditional form of one way communication. Furthermore, right now, the candidate that can uses online political tools, specifically video content, to mobilize their users by getting them to act offline, will be deemed most successful as constituents still value traditional forms of campaigning.

## 8. Reference

[1] Marin, Alexandra, and Barry Wellman. "Social Network Analysis: An Introduction." Computing in the Humanities and Social Sciences. Web. 04 Apr. 2010. <http://www.chass.utoronto.ca/~wellman/publications/newbies/newbies.pdf>.

[2] Oblak, Tanja. "Internet Kao Medij i Normalizacija Kibernetskog Prostora." Hrčak Portal Znanstvenih časopisa Republike Hrvatske. Web. 06 Apr. 2010. <hrcak.srce.hr/file/36810>.

[3] "Mixing Friends with Politics: A Functional Analysis of '08 Presidential Candidates Social Networking Profiles Authored by Compton, Jordan." All Academic Inc. (Abstract Management, Conference Management and Research Search Engine). Web. 13 June 2010. <http://www.allacademic.com//meta/p_mla_apa_research_citation/2/5/9/3/4/pages259348/p259348-1.php>.

[4] " All Academic Inc. (Abstract Management, Conference Management and Research Search Engine). Web. 13 June 2010. <http://www.allacademic.com//meta/p_mla_apa_research_citation/2/5/9/3/4/pages259348/p259348-1.php>.

[5] Собрание на Република Македонија. Web. 14 May 2010. <http://sobranie.mk/>.

[6] "President of the Republic of Macedonia." Web. 14 May 2010. <http://www.president.mk/en.html>.

[7] Government of Republic of Macedonia. Web. 15 May 2010. <http://www.vlada.mk/>.

[8] "Republic of Macedonia, Ministry of Foreign Affairs." Political Structure. Web. 19 June 2010. <http://www.mfa.gov.mk/>.

[9] ВМРО ДПМНЕ - официјална страница. Web. 17 May 2010. <http://www.vmro-dpmne.org.mk/>.

[10] Социјалдемократски Сојуз на Македонија. Web. 16 May 2010. <http://www.sdsm.org.mk/>.

[11] Bashkimi Demokratik Për Integrim. Web. 19 May 2010. <http://www.bdi.org.mk/>.

[12] "Alexa Internet - Website Information." Alexa the Web Information Company. Web. 20 June 2010. <http://www.alexa.com/siteinfo>.

[13] "What Is Web 2.0 - O'Reilly Media." O'Reilly Media - Technology Books, Tech Conferences, IT Courses, News. Web. 27 May 2010. <http://oreilly.com/web2/archive/what-is-web-20.html>.

[14] Web 2.0 Sites. Web. 28 May 2010. <http://web2.ajaxprojects.com/>.

[15] "Key Differences between Web1.0 and Web2.0." CiteSeerX. Web. 25 May 2010. <http://citeseerx.ist.psu.edu/viewdoc/summary?doi=10.1.1.145.3391>.

[16] "Political Participation and Web 2.0." Political Theory. Web. 24 May 2010. <http://arts.monash.edu.au/psi/news-and-events/apsa/refereed-papers/political-theory/allisonpoliticalparticipationandweb.pdf>.

[17] Државен Завод за Статистика на Македонија. Web. 28 May 2010. <http://www.stat.gov.mk/pdf/2009/8.1.9.23.pdf>.

[18] "Alexa - Top Sites in Macedonia." Alexa the Web Information Company. Web. 01 June 2010. <http://www.alexa.com/topsites/countries/MK>.

[19] "Statistics | Facebook." Welcome to Facebook. Web. 29 June 2010. <http://www.facebook.com/press/info.php?statistics>.

[20] Facebook Marketing Statistics, Demographics, Reports, and News – CheckFacebook. Web. 29 May 2010. <http://www.checkfacebook.com/>.

[21] "Statement of Rights and Responsibilities | Facebook." Welcome to Facebook. Web. 29 May 2010. <http://www.facebook.com/terms.php>.

[22] "What Is Social Networking? - Should You Join." What Is Social Networking? - What Is Social Networking? Web. 30 May 2010. <http://www.whatissocialnetworking.com/Should_You_Join.html>.

[23] Boyd, Danah M., and Nicole B. Ellison. "Social Network Sites: Definition, History". Web. 30 May 2010. <http://jcmc.indiana.edu/vol13/issue1/boyd.ellison.html>.

[24] "How Can Facebook Crack Its Advertising Problem?" Social Media News and Web Tips – Mashable – The Social Media Guide. Web. 21 May 2010. <http://mashable.com/2008/12/15/facebook-advertising-solution/>.

[25] Државна изборна комисија. Web. 01 June 2010. <http://www.sec.mk/>.

[26] Арсовски, Дамјан. "Најпопуларните друштвени мрежи во Македонија." IT.com.mk – Македонски ИТ портал. 13 Aug. 2009. Web. 05 June 2010. <http://it.com.mk/index.php/Statii/Veb/Najpopularnite-drushtveni-mrezhi-vo-Makedonija>.

[27] Арсовски, Виктор. "Facebook напредува со „брзина на светлината" IT.com.mk – Македонски ИТ портал. 19 Dec. 2009. Web. 11 June 2010. <http://it.com.mk/index.php/Vesti/Svet/Facebook-napreduva-so-brzina-na-svetlinata>.

[28] Булдиоски, Дарко. "Сите се на Facebook, или не? Комуникации." IT.com.mk – Македонски ИТ портал. 28 Jan. 2009. Web. 11 June 2010. <http://komunikacii.net/2009/01/28/facebook-users-balkan/>.

[29] Арсовски, Дамјан. "Facebook повторно го промени изгледот." IT.com.mk – Македонски ИТ портал. 23 Mar. 2009. Web. 11 June 2010. <http://www.it.com.mk/index.php/Vesti/Internet/Facebook-povtorno-so-promeni-vo-izgledot>.

[30] Арсовски, Дамјан. "Facebook со 200 милиони активни корисници." IT.com.mk – Македонски ИТ портал. 09 Apr. 2009. Web. 11 June 2010. <http://www.it.com.mk/index.php/Vesti/Internet/Facebook-povtorno-so-promeni-vo-izgledot>.

[31] YouTube - Broadcast Yourself. Web. 05 May 2010. <http://www.youtube.com/t/about>.

[32] "The YouTube Presidency." Blog Directory (washingtonpost.com). Web. 23 June 2010. <http://voices.washingtonpost.com/44/2008/11/the-youtube-presidency.html>.

**S. Emruli,** received his bachelor degree from Faculty of Communication Sciences and Technologies in Tetovo SEE University (2006), MSc degree from Faculty o f Organization and Informatics, Varaždin (2010). Currently works as professional IPA Advisor at Ministry of Local Self Government in Macedonia.

**T. Zejneli,** received his bachelor degree from Faculty of Communication Sciences and Technologies in Tetovo SEE University (2006). Currently works as Database administrator at Municipality of Tetovo in Macedonia.

**F. Agai,** received his bachelor degree from Faculty of Electronic Engineering in Skopje "St. Kiril and Metodij" University (1998), MSc degree from Faculty of Electronics, Telecommunication and Informatics in University of Pristina. Currently works as Professor at Electronics High School in Gostivar, Macedonia.